\documentclass[%
 reprint,
superscriptaddress,
 nofootinbib,
 amsmath,amssymb,
 aps,
 prd,
]{revtex4-2}

\usepackage{graphicx}
\usepackage{dcolumn}
\usepackage{bm}
\usepackage{hyperref}
\hypersetup{
    colorlinks=true,
    allcolors=blue
}

\usepackage{braket}
\usepackage{simpler-wick}
\usepackage{subfigure}
\usepackage[bb=dsserif]{mathalpha}
\usepackage{bm}

\newcommand{\fig}{Fig.\ }
\newcommand{\Fig}{Figure }
\newcommand{\sect}{Sec.\ }
\newcommand{\tab}{Table }

\newcommand{\ord}{\mathcal{O}}
\newcommand{\Op}{\mathcal{O}}
\newcommand{\dd}{\mathrm{d}}
\newcommand{\tr}{\operatorname{tr}}
\newcommand{\TT}{\operatorname{T}}

\newcommand{\LCS}{\sigma}
\newcommand{\im}{\operatorname{Im}}
\newcommand{\one}{\mathbb{1}}
\newcommand{\mvec}[1]{\vec{\mskip 0.5mu #1}\mskip 1.5mu} 

\newcommand{\red}[1]{#1}

\begin{document}


\title{Valence quark PDFs of the proton from two-current correlations in lattice QCD}

\author{Christian Zimmermann}
\affiliation{Aix Marseille Univ, Universit\'e de Toulon, CNRS, CPT, Marseille, France}%
\email{christian.zimmermann@univ-amu.fr}
\author{Andreas Sch\"afer}
\affiliation{Institute for Theoretical Physics, University of Regensburg, 93040 Regensburg, Germany}%
\email{andreas.schaefer@physik.uni-regensburg.de}

\date{\today}

\begin{abstract}
Following previous works on that topic, we consider Euclidean hadronic matrix elements in position space of two spatially separated local currents on the lattice, in order to extract the $x$ dependence of parton distribution functions (PDFs). The corresponding approach is often referred to by the term \textit{lattice cross section}. In this work we will consider valence quark PDFs of an unpolarized proton. We adapt the previously established formalism to our choice of operators. The calculation of the two-current matrix elements requires the evaluation of four-point functions. The corresponding calculation is carried out on a $n_f = 2+1$ gauge ensemble with lattice spacing $a = 0.0856~\mathrm{fm}$ and pseudoscalar masses $m_\pi = 355~\mathrm{MeV}$ and $m_K = 441~\mathrm{MeV}$. The four-point functions have been evaluated in a previous project. The lattice data is converted to the $\overline{\mathrm{MS}}$ scheme at a scale $\mu=2~\mathrm{GeV}$ and improved with respect to lattice artifacts. We use a common model as fit ansatz for the lattice data in order to extract the PDFs.
\end{abstract}

\maketitle


\section{\label{sec:intro}Introduction}

For decades, lattice calculations for quark parton distribution functions (PDFs) have continuously improved, a development which is documented, e.g. by the Lattice, the PDFlattice and other series of conferences and summary articles. A somewhat accidental selection is given by \cite{Lin:2017snn,Constantinou:2020hdm,Gross:2022hyw}. The classical approach of accessing PDFs on the lattice is the calculation of Mellin moments. The main problem in this context is that due to operator mixing one can at most calculate the leading three or four moments of each PDF, while fits like, e.g., the Hera legacy fit use already five parameters per quark flavor. Thus, computations of Mellin moments on the lattice can improve phenomenological PDF fits but not replace them. 

In the last decade, new methods for a direct access of the Bjorken-$x$ dependence on the lattice have been established. In particular, this includes the quasi-PDF approach in large momentum effective theory (LaMET) \cite{Ji:2013dva}, the Ioffe-time approach (pseudo-PDFs) \cite{Radyushkin:2016hsy}, or "OPE without OPE" \cite{Chambers:2017dov}. Some further pioneering publications on the PDF $x$ dependence on the lattice are given by \cite{Detmold:2005gg,Braun:2007wv}.

Notice that, for all these methods, there are certain complications. For instance, LaMET \cite{Ji:2013dva} requires to take the limit of very large hadron momentum $P_z$ but in practice, the rapid increase of the statistical uncertainty with $P_z$ makes it difficult to reach clearly perturbative scales. This also shows up as large power corrections for small and large $x$ which further complicate the task. Thus, it is probably best to combine the results of all established lattice methods with all phenomenological input. Over the last years, tremendous progress was made in obtaining relevant lattice results for all methods, far too much to give full credit to all investigations here. Therefore, we only cite a few recent publications on PDFs (the topic of this paper) in which many earlier references can be found: \cite{Delmar:2023agv, Alexandrou:2021oih,Bhat:2022zrw,Cichy:2021ewm,Gao:2022uhg,Gao:2023ktu,HadStruc:2022nay,Egerer:2021ymv,LatticeParton:2022xsd,Bringewatt:2020ixn,JeffersonLabAngularMomentumJAM:2022aix}. An overview of all important approaches regarding the PDF $x$ dependence on the lattice containing some state-of-the-art results is given by \cite{Constantinou:2020pek}.

The idea of a global analysis of the PDF lattice data motivates the introduction of the \textit{lattice cross section (LCS)}. This term refers to \red{a class of observables} that can be calculated on the lattice with a well-defined continuum limit and can be factorized in terms of PDFs and hard coefficients \cite{Ma:2014jla}. The latter is in analogy to the hadronic scattering tensor, which factorizes in terms PDFs and hard cross sections. For instance, quasi-PDFs are found to fulfill these conditions and, therefore, serve as LCS. Another class of suitable matrix elements is given by matrix elements of two spatially separated local quark currents \cite{Ma:2017pxb}. Calculations using that approach have already been performed in the past for the pion \cite{Sufian:2019bol,Sufian:2020vzb}. 

In the current work, we employ the LCS approach using two-current matrix elements to investigate valence quark PDFs for the unpolarized proton. The required matrix elements have been already generated in the context of our work on double parton distributions (DPDs) \cite{Diehl:2011yj,Bali:2020mij,Bali:2021gel,Reitinger:2024ulw} using a CLS gauge ensemble. CLS ensembles have been already used by RQCD to calculate Mellin moments \cite{Burger:2021knd,Bali:2023sdi} and full PDFs \cite{LatticeParton:2022xsd} using LaMET \cite{Ji:2020ect}, thus allowing to combine both sets of results.

This paper is organized as follows: In \sect\ref{sec:me} we review the relations between two-current matrix elements and PDFs as a consequence of the operator product expansion (OPE) and adapt the formalism to our choice of operators. The two-current matrix elements are obtained by calculating four-point functions. The corresponding simulations, as well as improvements with respect to lattice artifacts, are described in \sect\ref{sec:latt}. Our lattice results, as well as the extraction of the PDFs themselves, are described in \sect\ref{sec:res} before we conclude in \sect\ref{sec:concl}.

\vspace*{0.5cm}

\section{\label{sec:me}Matrix elements and PDFs}

In this section, we want to recall the relations between two-current matrix elements and PDFs following from the LCS approach, as first discussed by \cite{Ma:2014jla,Ma:2017pxb}. We consider a Lorentz-scalar hadronic matrix element of a nonlocal operator product $\Op_n(y)$, where $y$ denotes the maximal distance between the involved fields:

\begin{align}
\LCS_n^{(h)}(\omega,y^2) = \bra{h(p)} \TT\left\{ \Op_n(y) \right\} \ket{h(p)} \,,
\end{align}
where $\mathrm{T}$ denotes time ordering and $\omega = py$ is the so-called Ioffe time. According to the discussion in \cite{Ma:2014jla,Ma:2017pxb}, this kind of matrix element represent a suitable or "good" lattice cross section, if they can be consistently factorized in terms of PDFs and matching coefficients. Moreover, they need to be calculable on the lattice in Euclidean spacetime and have a well-defined continuum limit.

It has been shown in \cite{Ma:2014jla} that e.g.\ quasi-PDFs \cite{Ji:2013dva} represent a special class of LCSs in momentum space. Another class of suitable LCSs is given by matrix elements of two spatially separated quark currents:

\begin{align}
&\LCS^{(h)}_{ij,P}(\omega, y^2) := P_{\mu_1 \dots \mu_{n_i} \nu_1 \dots \nu_{n_j}} \nonumber\\
&\quad\times \left.\bra{h(p)} \TT\left\{ J_i^{\mu_1 \dots \mu_i}(y)\ J_j^{\nu_1 \dots \nu_j}(0) \right\} \ket{h(p)}\right|_{y^0 = 0} \,,
\label{eq:tcme-def}
\end{align}
with the local tensor-valued quark currents $J_i^{\mu_1\dots\mu_i}(y)$. These operators are understood to be renormalized in a suitable scheme. We use the Lorentz tensor $P$ to project on a scalar quantity, such that the lhs depends only on the Lorentz scalars $\omega = py$ and $y^2$.

The product of the two operators in \eqref{eq:tcme-def} can be expressed as a series of local operators by applying the OPE. Keeping only leading twist contributions, we can identify the PDFs $f_a^h(x,\mu^2)$:

\begin{align}
\LCS^{(h)}_{ij,P}(\omega, y^2)\! &:=\! \sum_a\! \int\! \frac{\dd x}{x} f^h_a(x,\mu^2)\ K_{ij,P}^{a}(x\omega,y^2,x^2,\mu^2) \nonumber\\
 &\ +\ord (y^2 \Lambda_{\mathrm{QCD}}^2) \,, \label{eq:tcme-ope}
\end{align}
where $K_{ij,P}^{a}(x\omega,y^2,x^2,\mu^2)$ are the so-called \textit{matching coefficients}, which can be determined perturbatively.

It has been shown in \cite{Ma:2017pxb} that \eqref{eq:tcme-ope} is valid for all $\omega$ and $y^2 p^2$, as long as $y$ is indeed a short distance as required by the OPE. Corrections contribute at $\Op(y^2 \Lambda_{\mathrm{QCD}}^2)$. Notice that this makes a calculation in position space mandatory since it is the only way to guarantee that $y^2 \Lambda_{\mathrm{QCD}}^2$ is in fact sufficiently small. A calculation in momentum space would always lead to contaminations from large distances $y$ \cite{Sufian:2019bol}. 

Comparing with other approaches, the advantage of the method described above is that it involves only local quark bilinears, which are renormalized multiplicatively in a well-known way. Moreover, the formulation is Lorentz covariant, so that we do not have to rely on large momentum scales. However, since we are restricted to small quark distances $y$, the use of large momenta is nevertheless advisable in order to cover a wide Ioffe time range.

The coefficients $K^{a}_{ij}$ in \eqref{eq:tcme-ope} only depend on the operators, i.e.\ they are independent of the external state. Hence, they can be determined by considering quark states. At leading order of $\alpha_s$, the PDF of a quark $q$ is $f_a^q(x,\mu^2) = \delta^q_a \delta(1-x)$ and \eqref{eq:tcme-def} becomes 

\begin{align}
\LCS^{(q)}_{ij}(\omega, y^2) = K_{ij}^{q}(\omega,y^2,x^2,\mu^2) + \ord (y^2 \Lambda_{\mathrm{QCD}}^2) \,,
\label{eq:tcme-ope-q-lo}
\end{align}
where $\LCS^{(q)}$ is the matrix element with respect to an external quark, which can be evaluated perturbatively.

Throughout this work, we restrict ourselves to local vector and axial vector currents:

\begin{align}
J_{i,qq^\prime}^\mu(y) = \bar{q}(y)\ \Gamma^\mu_i\ q^\prime(y) \,,
\label{eq:current-def}
\end{align}
where $\Gamma^\mu_i = \Gamma^\mu_{\mathrm{V}} = \gamma^\mu$ or $\Gamma^\mu_i = \Gamma^\mu_{\mathrm{A}} = \gamma^\mu\gamma_5$. The corresponding matching coefficients shall be determined in the following. To this end, we consider specifically the matrix elements:

\begin{align}
M^{(h)}_{q,ij} = \bra{h(p)} \TT\left\{ J_{i,qq^\prime}^\mu(y) J_{j,q^\prime q}^\nu(0) \right\} \ket{h(p)} \label{eq:Mhq-def} \,, \\
\widetilde{M}^{(h)}_{q,ij} = \bra{h(p)} \TT\left\{ J_{i,q^\prime q}^\mu(y) J_{j,qq^\prime}^\nu(0) \right\} \ket{h(p)} \label{eq:Mthq-def} \,,
\end{align}
where $q$ is the quark of interest. Basically, we are free to choose any quark flavor for $q^\prime$; the validity of the factorization formula \eqref{eq:tcme-ope} remains untouched. However, as we will see in \sect\ref{sec:latt}, it is advantageous to consider an (auxiliary) quark flavor that does not correspond to a valence quark of the considered hadron. 

Considering approximately massless quarks, the matrix element in \eqref{eq:Mhq-def} can be evaluated at tree level for $h=q$ and $q \neq q^\prime$ by a straightforward calculation:

\begin{align}
M^{(q)}_{q,ij}(\omega, y^2) &= \bra{q(p)} \wick{
\bar{q}(y)\ \Gamma_i \c1{q}^\prime(y)\ \c1{\bar{q}}^\prime(0)\ \Gamma_j q(0) 
}
\ket{q(p)} \nonumber \\
&= -\frac{i e^{i \omega}}{4 \pi^2 y^4} \tr\left\{ \Gamma_i y\!\!\!/ \Gamma_j p\!\!\!/ \right\} \,,
\label{eq:tcme-q-lo}
\end{align} 
and similar for the case of \eqref{eq:Mthq-def}:

\begin{align}
\widetilde{M}^{(q)}_{q,ij}(\omega, y^2) &= \bra{q(p)} 
\wick{
\c1{\bar{q}}^\prime(y)\ \Gamma_i q(y)\ \bar{q}(0)\ \Gamma_j \c1{q}^\prime(0) 
}
\ket{q(p)} \nonumber\\
&= \frac{i e^{-i \omega}}{4 \pi^2 y^4} \tr\left\{ \Gamma_i y\!\!\!/ \Gamma_j p\!\!\!/ \right\} \,.
\label{eq:tcmet-q-lo}
\end{align}
At tree level, there is no factorization scale and we will drop the corresponding argument $\mu$ of the PDF and the matching coefficients in the following.

In the context of this work, we consider the product of two vector currents or two axial vector currents, i.e.\ $(\Gamma_i,\Gamma_j) = (\Gamma^\mu_{\mathrm{V}},\Gamma^\nu_{\mathrm{V}})$, or $(\Gamma_i,\Gamma_j) = (\Gamma^\mu_{\mathrm{A}},\Gamma^\nu_{\mathrm{A}})$, respectively. In the end, we analyze the average of hadronic matrix elements of both types of operator combinations in order to suppress lattice artifacts as will be explained in \sect\ref{sec:latt}:

\begin{align}
M_q^{(h)\mu\nu}(p,y) = {\textstyle\frac{1}{2}} \left[ M_{q,VV}^{(h)\mu\nu}(p,y) + M_{q,AA}^{(h)\mu\nu}(p,y) \right] \,.
\label{eq:tcme-VV-AA}
\end{align}
In both cases, the traces in the expressions for the previously derived matching coefficients \eqref{eq:tcme-q-lo} and \eqref{eq:tcmet-q-lo} for quarks give

\begin{align}
\tr\left\{ \gamma^{\mu} y\!\!\!/ \gamma^{\nu} p\!\!\!/ \right\} 
&= \tr\left\{ \gamma^{\mu} \gamma_5 y\!\!\!/ \gamma^{\nu} \gamma_5 p\!\!\!/ \right\} 
\nonumber\\
&= 4 \left( p^{\mu} y^{\nu} + p^{\nu} y^{\mu} - g^{\mu\nu} \omega \right) \,.
\label{eq:tcme-q-lo-trace}
\end{align}
Hence, in order to extract the PDF from the matrix element \eqref{eq:tcme-VV-AA}, we have to project out the part that is proportional to $( p^{\mu} y^{\nu} + p^{\nu} y^{\mu} - g^{\mu\nu} \omega )$. To this end, we symmetrize $M^{\mu\nu}$ with respect to $\mu$ and $\nu$ and decompose it in terms of Lorentz invariant functions and suitable basis tensors:

\begin{align}
&y^4 M_q^{(h)\{\mu\nu\}}(p,y) = p^{\mu} p^{\nu} A^h_q(\omega, y^2) \nonumber\\ 
&\quad + m^2 g^{\mu\nu} B^h_q(\omega, y^2) + m^4 y^{\mu} y^{\nu} C^h_q(\omega, y^2) \nonumber\\
&\quad + m^2 \left( p^{\mu} y^{\nu} + p^{\nu} y^{\mu} - g^{\mu\nu} \omega \right) D^h_q(\omega, y^2) \,,
\label{eq:tcme-VV-AA-decomp}
\end{align}
%
where braces indicate normalized symmetrization, i.e.\ $M^{\{\mu\nu\}} = \frac{1}{2}\left[ M^{\mu\nu} + M^{\nu\mu} \right]$. The purpose of the factor $y^4$ on the lhs will become clear later. The desired information is encoded in the invariant function $D(\omega, y^2)$. It can be obtained by applying the following projector to the hadronic two-current matrix element:

\begin{widetext}
\begin{align}
P_{D}^{\mu\nu}(p,y) := 
\frac{
\omega \left( m^2 y^2 - \omega^2 \right) g^{\mu\nu} + \left( m^2 y^2 + 2\omega^2 \right)\ \left( p^{\mu} y^{\nu} + p^{\nu} y^{\mu} \right) - 3 \omega m^2 y^{\mu} y^{\nu} - 3 \omega y^2 p^{\mu} p^{\nu}
}{
2\left( \omega^2 - m^2 y^2 \right)^2
} \,,
\end{align}
\end{widetext}
so that

\begin{align}
P_{D,\mu\nu}(p,y)\ M_q^{(h)\{\mu\nu\}}(p,y) &= \frac{m^2}{y^4} D^h_q(\omega, y^2)
\label{eq:proj-tcme-d}
\end{align}
Applying the same projector to the trace \eqref{eq:tcme-q-lo-trace} in the tree-level expression, we obtain

\begin{align}
&P_{D,\mu\nu}(p,y)\ \tr\left\{ \gamma^{\mu} y\!\!\!/ \gamma^{\nu} p\!\!\!/ \right\} \nonumber \\
&\quad = P_{D,\mu\nu}(p,y)\ \tr\left\{ \gamma^{\mu} \gamma_5 y\!\!\!/ \gamma^{\nu} \gamma_5 p\!\!\!/ \right\} = 4 \,.
\label{eq:proj-tcme-q-d}
\end{align}
Combining the Eqs.\ \eqref{eq:tcme-q-lo}, \eqref{eq:tcme-q-lo-trace}, \eqref{eq:proj-tcme-d} and \eqref{eq:proj-tcme-q-d}, we find

\begin{align}
&m^2 D^h_q(\omega,y^2) = y^4 P_{D,\mu\nu}(p,y)\ M_q^{(h)\{\mu\nu\}}(p,y) \nonumber\\
&\quad = y^4 \sigma^{(h)}_{q,P_{D}}(\omega,y^2) = -\frac{i}{\pi^2} \int \dd x\ f^h_q(x)\ e^{i x\omega} \,.
\end{align}
Since the rhs does not depend on $y^2$, we expect the same to be true for the Lorentz invariant function $D^h_q(\omega,y^2)$, as long as we are in the kinematic region where our OPE approach is valid and assuming that higher-twist effects and higher-order corrections are negligible. Nevertheless, we will keep the argument $y^2$ for the moment, since a dependence on $y^2$ is considered technically in our analysis. In the end, we obtain the relation between the invariant function $D^h_q(\omega,y^2)$ and the desired PDF $f^h_q(x)$:

\begin{align}
D^h_q(\omega,y^2) = -\frac{i}{\pi^2 m^2} \int \dd x\ f^h_q(x)\ e^{i x\omega} \,.
\label{eq:master}
\end{align}
In the following, we only consider the proton, i.e.\ $h=p$, and drop the corresponding superscripts for the matrix elements and invariant functions.

\section{\label{sec:latt}Lattice simulation}

\begin{figure*}
\includegraphics[scale=1]{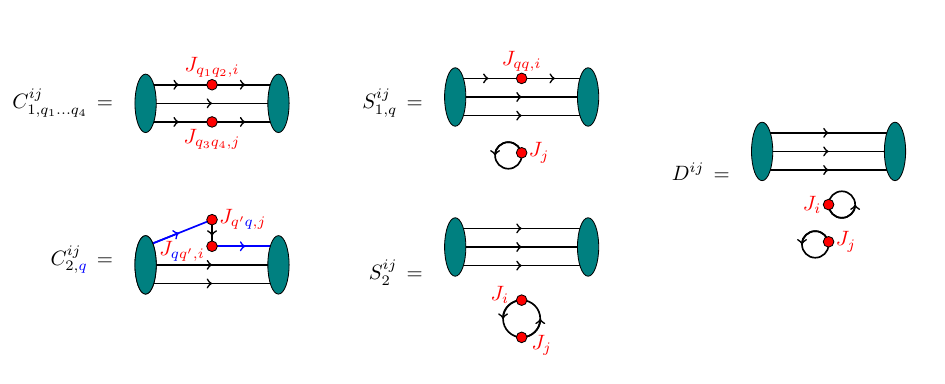}
\caption{All types of Wick contractions contributing to $C_{4\mathrm{pt}}$. In the case of the flavor combination required in the context of this work, only the $C_2$ contraction has to be evaluated. For this graph, $q$ indicates the quark flavor of the quark lines connected to the source and the sink. The corresponding quark lines are shown in blue color. \label{fig:wick}}
\end{figure*}

The evaluation of the two-current matrix element in \eqref{eq:Mhq-def} and \eqref{eq:Mthq-def} requires the calculation of a four-point function. In this work, we reuse the data that have been already generated in the context of \cite{Bali:2021gel}. In the following, we give a brief overview of the corresponding lattice techniques. 

\subsection{\label{ssec:wick}Four-point functions and Wick contractions}

A hadronic two-current matrix element can be evaluated directly on the lattice if the two currents are located at equal time. The unpolarized matrix element for the proton can be expressed in terms of four-point functions $C_{4\mathrm{pt}}(t,\tau,\mvec{y})$ in the limit of large time separations, where excited states are expected to be suppressed:

\begin{align}
\bra{p} J(\mvec{y}) J(\mvec{0}) \ket{p} &= 2V E_{\mvec{p}} \left.\frac{C_{4\mathrm{pt}}(t,\tau,\mvec{y})}{C_{2\mathrm{pt}}(t)}\right|_{0\ll \tau \ll t} \,,
\label{eq:tcme-4pt}
\end{align}
where $E_{\mvec{p}} := \sqrt{m^2+\mvec{p}}$ and $V$ is the spatial lattice volume. The four-point function and two-point function are defined, respectively, as

\begin{align}
&C_{4\mathrm{pt}}(t,\tau,\mvec{y}) 
:= 
a^6 \sum_{\mvec{z},\mvec{z}^\prime} e^{-i\mvec{p}(\mvec{z}^\prime-\mvec{z})}
\nonumber\\
&\quad\times
\left\langle \tr\left\{ 
P_+ \mathcal{P}(\mvec{z}^\prime,t) J_i(\mvec{y},\tau) J_j(\mvec{0},\tau) \overline{\mathcal{P}}(\mvec{z},0) 
\right\} \right\rangle \,, \label{eq:4pt-def} \\
&C_{2\mathrm{pt}}(t) 
:= 
a^6 \sum_{\mvec{z},\mvec{z}^\prime} e^{-i\mvec{p}(\mvec{z}^\prime-\mvec{z})} 
\nonumber\\
&\quad\times
\left\langle \tr\left\{ 
P_+ \mathcal{P}(\mvec{z}^\prime,t) \overline{\mathcal{P}}(\mvec{z},0) 
\right\} \right\rangle \,,
\label{eq:2pt-def}
\end{align}
with the parity projector

\begin{align}
P_+ := {\textstyle\frac{1}{2}} \left( \one + \gamma_4 \right)
\end{align}
and the proton creation and annihilation operators

\begin{align}
\overline{\mathcal{P}}(\mvec{x},t) 
&:= 
\left.\epsilon_{abc}
\left[ 
\bar{u}_a(x)\ C\gamma_5\ \bar{d}_b^{T}(x)
\right] \bar{u}_c(x) \right|_{x^4=t} \,,
\nonumber\\
\mathcal{P}(\mvec{x},t) 
&:= 
\left.\epsilon_{abc}\ 
u_a(x) \left[ 
u_b^T(x)\ C\gamma_5\ d_c(x)
\right]\right|_{x^4=t} \,.
\end{align}
The matrix elements are understood to be renormalized and converted to the $\overline{\mathrm{MS}}$ scheme for a scale $\mu = 2~\mathrm{GeV}$. Here we use the renormalization factors $Z_V = 0.7128$ and $Z_A = 0.7525$ \cite{RQCD:2020kuu}.

The four-point function \eqref{eq:4pt-def} decomposes into several Wick contractions. For the proton, there are in general five types, which are shown in \fig\ref{fig:wick}.
In \cite{Bali:2021gel}, \sect 3.2, explicit expressions for the contractions for all relevant flavor combinations have been given and the required evaluation methods have been discussed in detail. The exact contribution of Wick contractions depends on the quark flavors of the currents. Considering \eqref{eq:Mhq-def} and \eqref{eq:Mthq-def}, $q$ denotes the quark flavor that corresponds to the quark of the considered PDF $f_q$. As described in \sect\ref{sec:me}, the factorization \eqref{eq:tcme-ope} can be performed for any quark flavor $q^\prime$; all choices are suitable for treating the PDF $f_q$. However, we can simplify the calculation by choosing a quark flavor for which the number of contributing Wick contractions is minimal. If we consider $q \neq q^\prime$, the diagrams $S_1$ and $D$ do not contribute. Moreover, if $q^\prime$ is not a valence quark, there is no contribution from the $C_1$ contraction. Hence, we are left with the contractions $C_{2}$ and $S_2$, if we choose $q^\prime$ to be a nonvalence quark that is different from the quark of interest $q$. Notice that the explicit contribution of contractions with $C_{2}$ topology again depends on the quark flavor $q$, which is why it is indicated in the subscript, i.e.\ $C_{2,q}$. The corresponding quark lines are shown in blue in \fig\ref{fig:wick}. As long as $q^\prime$ has the same mass as $q$, there is no dependence on the flavor $q^\prime$. Explicitly, we find the following contribution for the matrix elements \eqref{eq:Mhq-def} and \eqref{eq:Mthq-def}:

\begin{align}
M_{q,ij}(p,y) &= C_{2,q}^{ij}(p,y) + S_{2}^{ij}(p,y) \nonumber \\
\widetilde{M}_{q,ij}(p,y) &= C_{2,q}^{ji}(p,-y) + S_{2}^{ij}(p,y) \,.
\end{align}
The situation simplifies drastically if we consider valence quark PDFs $f_{q,v}$. In this case, we have to calculate the difference of $M$ and $\widetilde{M}$. For this combination, the disconnected diagram $S_2$ cancels exactly, so that we only have to consider the connected contribution $C_{2,q}$ for the valence quark PDF $f_{q,v}$:

\begin{align}
\left[ M_{q,ij}(p,y) - \widetilde{M}_{q,ij}(p,y) \right] &= 2i \im\left\{ C_{2,q}^{ij}(p,y) \right\} \,,
\end{align}
so that we obtain the relation

\begin{align}
\im \left\{ D_q(\omega,y^2) \right\} = -\frac{1}{\pi^2 m^2} \int_{0}^{1} \dd x\ f^p_{q,v}(x)\ \cos(x\omega) \,.
\label{eq:im-D-integ}
\end{align}
As a consequence of the number sum rule for PDFs, we find for $\omega = 0$:

\begin{align}
-\pi^2 m^2 \im \left\{ D_q(0,y^2) \right\} = \int_{0}^{1} \dd x\ f^p_{q,v}(x) = N_q \,,
\label{eq:number-sum-rule}
\end{align}
where $N_q$ is the number of valence quarks of flavor $q$ in the nucleon. Thus, it is useful to define:

\begin{align}
\widehat{D}_q(\omega) := -\frac{\pi^2 m^2}{N_q} D_q(\omega,y^2) \,.
\label{eq:Dhat-def}
\end{align}

\subsection{\label{ssec:aniso}Anisotropy reduction}

As already discussed in \cite{Bali:2021gel}, the $C_2$ contraction exhibits a strongly anisotropic behavior. This is mostly due to large lattice artifacts of the Wilson quark propagator $M_{\mathrm{latt}}(y)$ and has been already investigated in \cite{Bali:2018spj}. Following the idea of \cite{Bali:2018spj}, we implement a method to reduce lattice artifacts introduced by the quark propagator in the $C_2$ data.

First of all, we choose operator combinations where the chiral-odd part ($\propto \one$) of the propagator cancels exactly in leading-order perturbation theory. In the continuum, the propagator is dominated by its chiral-even part ($\propto y\!\!\!/$, for source-sink distance $y$), whereas the chiral-odd part is very small. This is different for the employed Wilson fermions due to the Wilson term, which suppresses the doublers. In this case, a suitable operator combination canceling the chiral-odd part is given by $VV+AA$, which justifies the choice of matrix elements in \eqref{eq:tcme-VV-AA}.

\begin{figure}
\vspace*{0.5cm}
\includegraphics[scale=0.45, clip, trim=0 0 0 1.8cm]{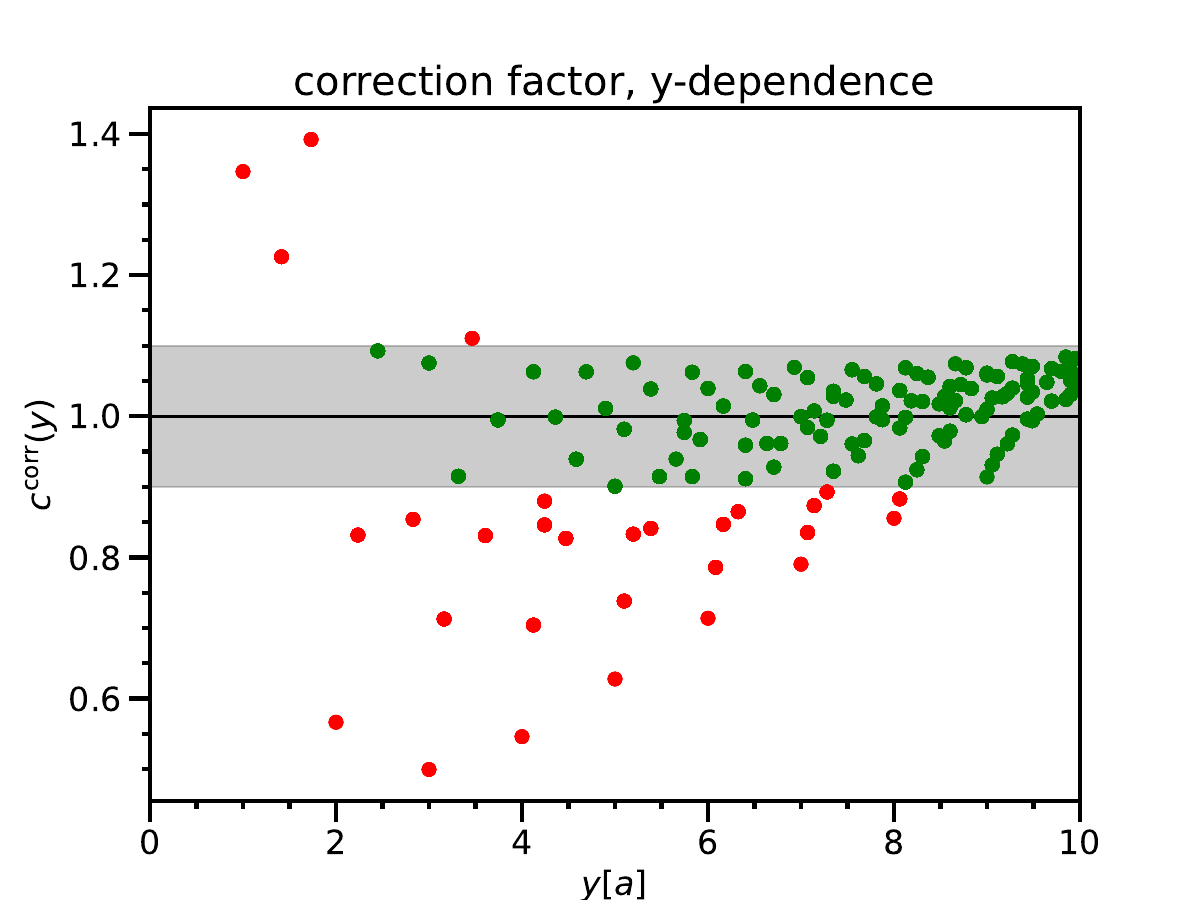}
\caption{Correction factor $c^{\mathrm{corr}}$ as a function of $y = |\mvec{y}|$, where each data point corresponds to a distance vector $\mvec{y}$. The gray band indicates a correction of at most $10\%$. Data points with larger corrections (red) are dropped. All other data points (green) are multiplied by $c^{\mathrm{corr}}(y)$. \label{fig:c_corr}}
\end{figure}

In order to deal with the anisotropy effects in the chiral-even part, we consider the quark propagator in the continuum $M_{\mathrm{cont}}(y)$ and define a correction factor $c^{\mathrm{corr}}(y)$ by

\begin{align}
\tr\left\{ y\!\!\!/ M_{\mathrm{cont}}(y) \right\} = c^{\mathrm{corr}}(y) \tr\left\{  M_{\mathrm{latt}}(y) \right\} \,.
\end{align}
At tree level, the correction factor can be expressed as

\begin{align}
c^{\mathrm{corr}}(y) = -\frac{m^2}{\pi^2} \frac{K_2(m\sqrt{-y^2}) }{ \tr\left\{ y\!\!\!/ M^{\mathrm{free}}_{\mathrm{latt}}(y) \right\} } \,,
\label{eq:c_corr_def}
\end{align}
where $M^{\mathrm{free}}_{\mathrm{latt}}$ is the free Wilson propagator and $K_2$ denotes the modified Bessel function. \Fig\ref{fig:c_corr} shows the values of $c^{\mathrm{corr}}$ for all data points in the plotted region of $y$. The jumps depend on the direction of $\mvec{y}$ and clearly indicate the anisotropy of the Wilson propagator.

For the subsequent analysis, we will keep only data points for which

\begin{align}
|c^{\mathrm{corr}}(y)-1| < 0.1 \,,
\end{align}
which is indicated by the gray band in \fig\ref{fig:c_corr}. The corresponding data points are colored green and the dropped data points red. Moreover, the remaining data points are multiplied by the correction factor

\begin{align}
C^{\mathrm{corr}}_{2,q}(y) = c^{\mathrm{corr}}(y)\ C_{2,q}(y) \,.
\end{align}

\subsection{\label{ssec:r}Lattice setup}

\begin{table*}
\begin{center}
\begin{tabular}{cccccccccccc}
\hline
\hline
ID\ &\ $\beta$\ &\ $a~[\mathrm{fm}]$\ &\ $L^3 \times T$\ &\ $\kappa_{l}$\ &\ $\kappa_{s}$\ &\ $m_{\pi}~[\mathrm{MeV}]$\ &\ $m_{K}~[\mathrm{MeV}]$\ &\ $m_\pi L a$\ &\ Configs \\
\hline
H102\ &\ $3.4$\ &\ $0.0856$\ &\ $32^3 \times 96$\ &\ $0.136865$\ &\ $0.136549339$\ &\ $355$\ &\ $441$\ &\ $4.9$\ &\ $2037$ \\
\hline
\hline
\end{tabular}
\end{center}
\caption{Parameters of the employed gauge ensemble, which has been generated by the CLS Collaboration \cite{Bruno:2014jqa}. In the present simulation, 990 configurations are used.\label{tab:lattice-info}}
\end{table*}

In our simulation we use a $32^3\times 96$ ensemble with open boundary conditions in time direction and pseudoscalar masses $m_\pi = 355~\mathrm{MeV}$ and $m_K = 441~\mathrm{MeV}$ generated by the CLS Collaboration \cite{Bruno:2014jqa}. It employs the tree-level improved L\"uscher-Weisz gauge action and $n_f = 2+1$ Sheikholeslami-Wohlert fermions. The parameters of the ensemble are summarized in \tab\ref{tab:lattice-info}. From this ensemble we use 990 configurations. Our approach requires data points within a wide range of $\omega$ and, at the same time, we have to keep the distance between the currents as small as possible in order to fulfill $y \ll \Lambda_{\mathrm{QCD}}^{-1}$. Hence, we evaluate the four-point function for the relatively high lattice momenta $\mvec{P} = (-2,-2,-2)$, $(2,2,-2)$, $(2,-2,2)$, $(-2,2,2)$; i.e.\ the absolute value of the physical momenta

\begin{align}
\mvec{p} = \frac{2\pi \mvec{P}}{La} \,
\end{align}
is $|\mvec{p}| \approx 1.57~\mathrm{GeV}$. Moreover, we also employ $P=(0,0,0)$ and $P=(-1,-1,-1)$ for consistency checks. The quark sources and sinks are improved by using momentum smearing \cite{Bali:2016lva} with $n=250$ smearing iterations. The nucleon mass is determined from the two-point function \eqref{eq:2pt-def} to be $m_N = 1.1296(75)~\mathrm{GeV}$.

In order to avoid artifacts due to the open boundary conditions in time direction, we place the nucleon source at $t_{\mathrm{src}} = T/2$, where $T$ is the lattice extension in time direction. The source-sink separation is chosen to be $t=t_{\mathrm{sink}}-t_{\mathrm{src}} = 12a$ for $\mvec{p} = \mvec{0}$ and $t = 10a$ for nonzero momentum. The $C_2$ contraction is evaluated for the insertion time $\tau = t_{\mathrm{ins}} - t_{\mathrm{src}} = t/2$, i.e.\ $\tau = 6a$ or $\tau = 5a$, depending on the momentum.

\section{\label{sec:res}Results}

\subsection{\label{ssec:res}Lattice data}

The data for the two-current matrix elements have been generated in the context of the study described in \cite{Bali:2021gel}. Therein, it was found that the four-point function data have good quality and a statistical error of several percent. Moreover, no evidence for significant excited state contributions has been found. As already mentioned, the data of the $C_2$ contraction, which is the only contraction that contributes in the present analysis, exhibit strong anisotropy effects. In the previous section, we described a method to deal with this complication. In the following, we consider the results obtained from the $C_2$ data, which have been improved according to the discussion in \sect\ref{ssec:aniso}. In order to obtain the desired Lorentz invariant functions, the system of equations given by \eqref{eq:tcme-VV-AA-decomp} is solved numerically for each value of $|\mvec{p}|$ taking into account all contributing nucleon momenta. 

\begin{figure}
\vspace*{0.5cm}
\subfigure[$D_u$, $y$-dependence at $\omega = 0$\label{fig:imD-u-ydep}]
{\includegraphics[scale=0.45, clip, trim=0 0 0 1.6cm]{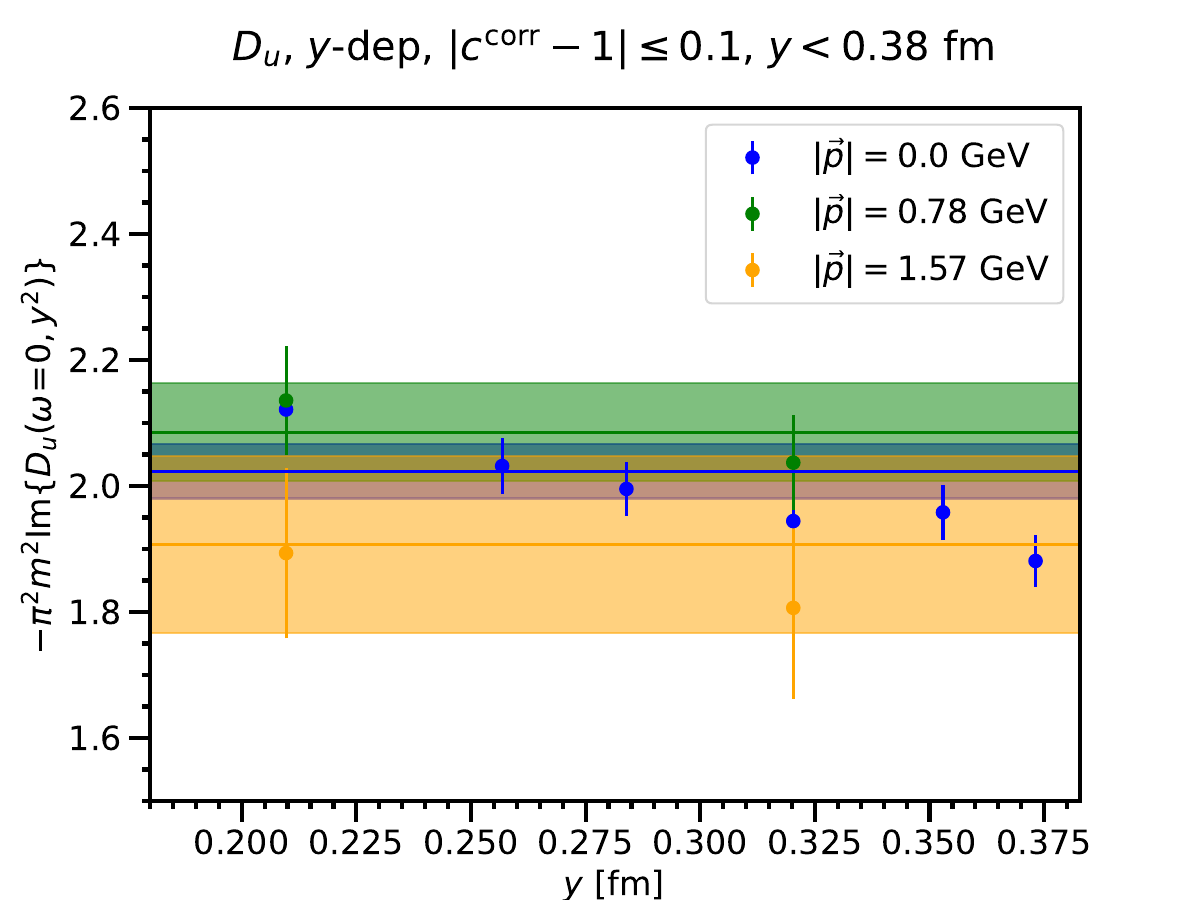}}
\subfigure[$D_d$, $y$-dependence at $\omega = 0$\label{fig:imD-d-ydep}]
{\includegraphics[scale=0.45, clip, trim=0 0 0 1.6cm]{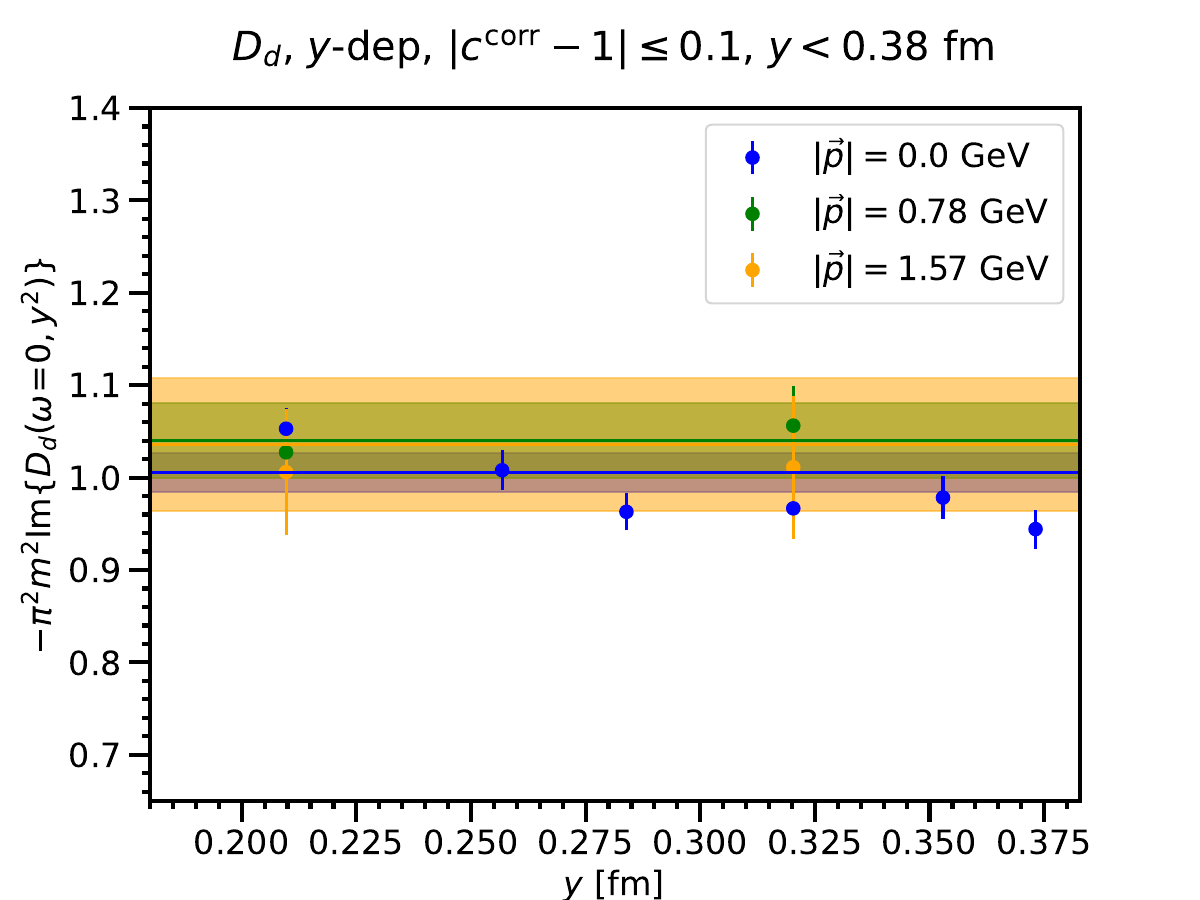}}
\caption{$y$ dependence of $D_q$ for quark flavors $u$ (a) and $d$ (b), where we compare contributions by momenta of absolute values $|\mvec{p}| = 0$ (blue), $|\mvec{p}| = 0.78~\mathrm{GeV}$ (green) and $|\mvec{p}| = 1.57~\mathrm{GeV}$ (orange). The bands show the results obtained by a solution of the system of equations \eqref{eq:tcme-VV-AA-decomp} where the $y$ dependence is neglected. \label{fig:imD-ydep}}
\end{figure}
\Fig\ref{fig:imD-ydep} shows the data for the lhs of \eqref{eq:number-sum-rule} ($\omega = 0$) for flavors $u$ (a) and $d$ (b) as a function of the operator distance $y=|\mvec{y}|$ for each $|\mvec{p}|$. The data is plotted for $|\mvec{y}| < 0.38~\mathrm{fm}$ ($|\mvec{y}|/a < 4.5$). In this region, we find that the results are consistent with the number sum rule \eqref{eq:number-sum-rule} within the statistical error. For $\mvec{p} = \mvec{0}$, where the errors are smallest, we can observe that the values are slightly decreasing for increasing $|\mvec{y}|$, which is in contrast to the prediction given by \eqref{eq:im-D-integ}. Let us recall that \eqref{eq:im-D-integ} was derived at leading order and leading twist only. Taking into account higher-order corrections may help to improve the situation. This will be considered in future works.

For further analysis steps, we solve again the system of equations \eqref{eq:tcme-VV-AA-decomp} taking into account all data points for $|\mvec{y}| < 0.38~\mathrm{fm}$ and neglecting the dependence of the invariant functions on $y^2$ (we indicate this by omitting the corresponding argument of $D$). The corresponding results for $\omega=0$ are represented by the bands in \fig\ref{fig:imD-ydep}. Moreover, we calculate the normalized invariant function $\widehat{D}_q$ defined in \eqref{eq:Dhat-def} by evaluating the ratio 

\begin{align}
\widehat{D}_q(\omega) = \frac{D_q(\omega)}{\im\left\{D_q(0)\right\}} \,.
\end{align}
The correlation between numerator and the denominator, leads to smaller statistical errors on $\widehat{D}_q$ compared to the non-normalized quantities $D_{q}$.
 
\begin{figure}
\vspace*{0.5cm}
\includegraphics[scale=0.43, clip, trim=0 0 0 1.8cm]{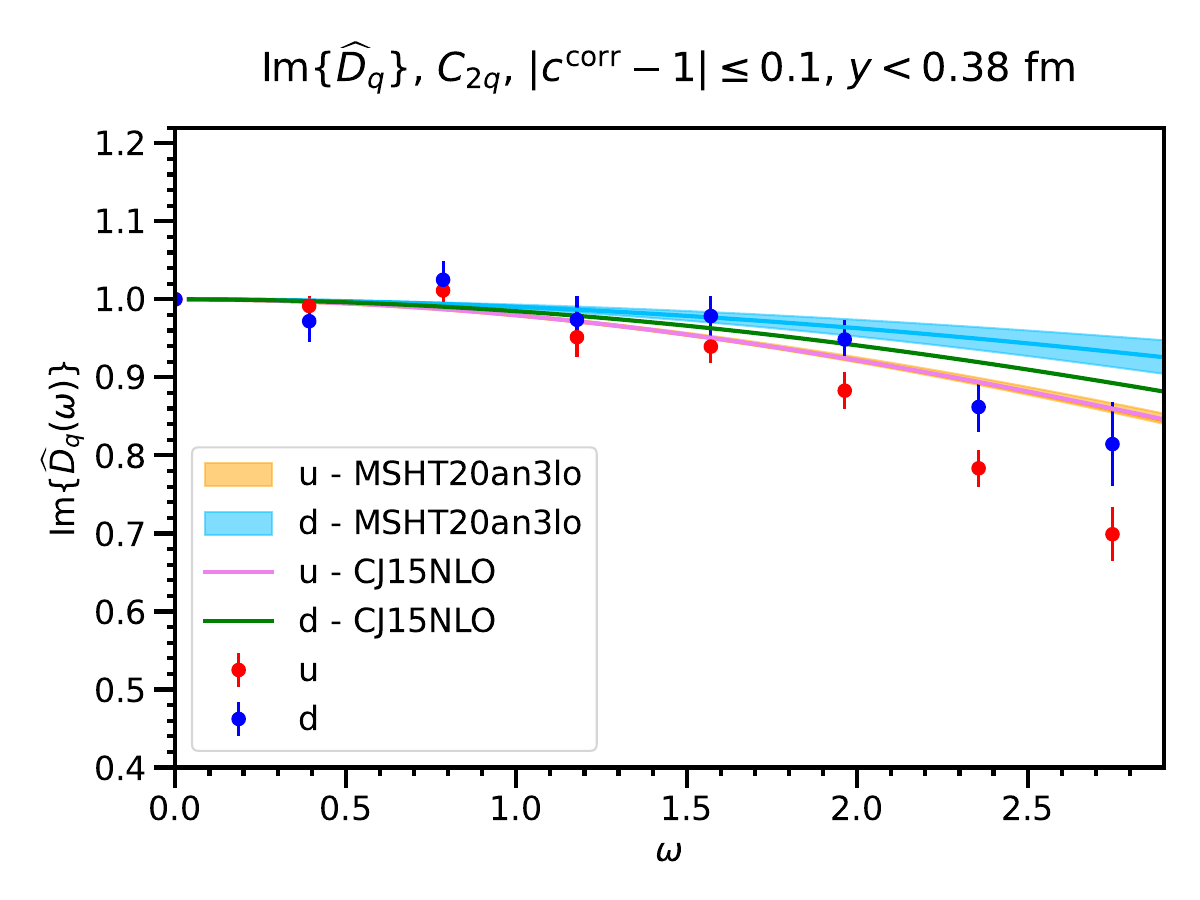}
\caption{Imaginary part of the Lorentz invariant function $\widehat{D}$ obtained from our lattice simulation for $|\mvec{p}|=1.57~\mathrm{GeV}$. This is shown for the $u$ quark (red) and the $d$ quark (blue) as a function of Ioffe time $\omega = py$. We also show the corresponding results obtained from experiments \cite{Accardi:2016qay} (purple, $u$; green, $d$) and \cite{McGowan:2022nag} (orange, $u$; light blue, $d$). \label{fig:imDhat-omega}}
\end{figure}
The corresponding result is plotted as a function of Ioffe time $\omega$ in \fig\ref{fig:imDhat-omega} for both quark flavors. In this figure, we also compare with results obtained from selected experimental data: This includes the datasets \cite{Accardi:2016qay}\footnote{The error bands for these datasets are not shown.}, where the corresponding analysis is in particular optimized for the large $x$ region, which we are interested in. Moreover, we compare with the datasets \cite{McGowan:2022nag}, which represents an example of a recent analysis. Both datasets are evolved to an evolution scale of $\mu = 2~\mathrm{GeV}$. The curves shown in the plot are obtained by inverting the Fourier transform in \eqref{eq:im-D-integ}. Comparing with our lattice results, we can observe discrepancies for larger values of $\omega$ starting at $\omega = 2$. Especially for quark flavor $u$, these differences tend to be large. Notice that data points for large values of $\omega$ correspond to large values of $|\mvec{y}|$. Taking into account higher-order corrections might reduce the differences. Another potential source of discrepancies are higher-twist contributions, which have been neglected in the derivation of the factorization formula \eqref{eq:tcme-ope}.

\subsection{\label{ssec:fits}Extraction of PDFs}

The invariant functions $\widehat{D}_q(\omega)$ are directly related to the PDFs $f^p_q(x)$ according to Eq.\ \eqref{eq:im-D-integ}. Because of the restriction of the accessible Ioffe time region and the fact that we can sample only a few points in that region, the inversion of \eqref{eq:im-D-integ} is highly ill posed. This is often referred to as the \textit{inverse problem}. In the recent past, several formalisms and methods have been developed in order to deal with that issue, like the Backus-Gilbert method, neural networks and Bayesian analysis methods \cite{Karpie:2019eiq}. In this first study for the case of the nucleon, we restrict ourselves to a fit to a function based on a commonly used model for the PDFs. This has already been applied for the calculations on pion PDFs \cite{Sufian:2019bol,Sufian:2020vzb}. We consider the following ansatz:

\begin{align}
\frac{f^p_q(x)}{N_q} 
&= 
N(\alpha,\beta,\rho,\gamma)\ 
x^\alpha (1-x)^\beta (1+\rho \sqrt{x}+\gamma x) \,,\nonumber\\
N(\alpha,\beta,\rho,\gamma) 
&:= \left[
B(1+\beta,1+\alpha) + \rho B(1+\beta,\textstyle{\frac{3}{2}}+\alpha)\right.\nonumber\\
&\left.+ \gamma B(1+\beta,2+\alpha) \right]\,, 
\label{eq:pdf-ansatz}
\end{align}
where $N_q$ is the number of quarks $q$ in the nucleon and $B$ is the Euler beta function. Our normalization is chosen so that the integral over $x$ equals $1$ by definition. The ansatz \eqref{eq:pdf-ansatz} is inserted into \eqref{eq:im-D-integ} yielding a function that can be used to fit the lattice data for $\im\{\widehat{D}_q\}$, where $\alpha$, $\beta$, $\rho$ and $\gamma$ have to be determined. The range of accessible data points with respect to Ioffe time $\omega$ only allows us to perform a two-parameter fit. Hence, we treat $\alpha$ and $\beta$ as free fit parameters with the phenomenologically motivated bounds

\begin{align}
-1 < \alpha < 0 \,, \qquad 0 < \beta \,.
\end{align}

The fits are performed for several fixed values for $\rho$ and $\gamma$ considering either $\rho$ or $\gamma$ to be zero and $\gamma,\rho < 10$. It turns out that the parameter $\alpha$, which determines the small-$x$ behavior, always tends to be zero. Since the small-$x$ region is likely to be governed by input from the region of large $\omega$, where we do not have any data points, we will exclude the results for $\alpha$ from our discussion. The results for $\beta$ and the corresponding $\chi^2/\mathrm{d.o.f.}$ for selected combinations of $\gamma$ and $\rho$ are compiled in \tab\ref{tab:omega-fit-results}.

\begin{table}
\begin{center}
\begin{tabular}{c|cc|c|c}
\hline
\hline
Flavor & $\gamma$ & $\rho$ & $\beta$ & $\chi^2/\mathrm{d.o.f.}$ \\
\hline
$u$ & $0$ & $0$ & $2.56(33)$ & $1.87$ \\
 & $5$ & $0$ & $3.86(40)$ & $1.80$ \\
 & $0$ & $5$ & $3.49(40)$ & $1.81$ \\
\hline
$d$ & $0$ & $0$ & $4.5(1.1)$ & $1.45$ \\
 & $5$ & $0$ & $6.2(1.3)$ & $1.43$ \\
 & $0$ & $5$ & $5.9(1.3)$ & $1.43$ \\
\hline
\hline
\end{tabular}
\end{center}
\caption{Fit results for the parameter $\beta$ and the $\chi^2/\mathrm{d.o.f.}$ for all flavors and selected combinations of $\gamma$ and $\rho$.\label{tab:omega-fit-results}}
\end{table}
\begin{figure}
\vspace*{0.5cm}
\subfigure[$\mathrm{Im}\{\widehat{D}_{u}(\omega)\}$, fit comparison \label{fig:imDhat-u-omega-fit-grcomp}]
{\includegraphics[scale=0.43, clip, trim=0.5cm 0 0 1.6cm]{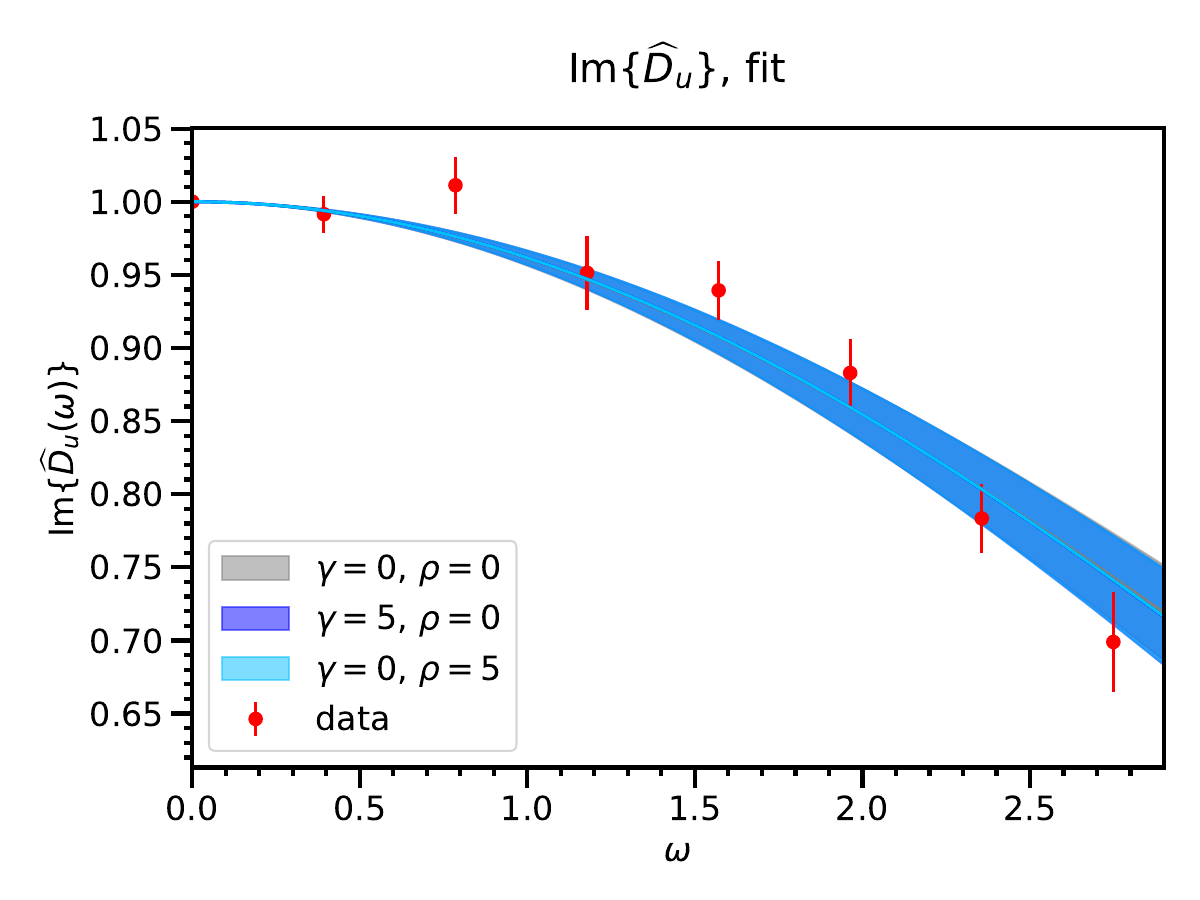}}
\vspace*{0.5cm}
\subfigure[$\mathrm{Im}\{\widehat{D}_{d}(\omega)\}$, fit comparison \label{fig:imDhat-d-omega-fit-grcomp}]
{\includegraphics[scale=0.43, clip, trim=0.5cm 0 0 1.6cm]{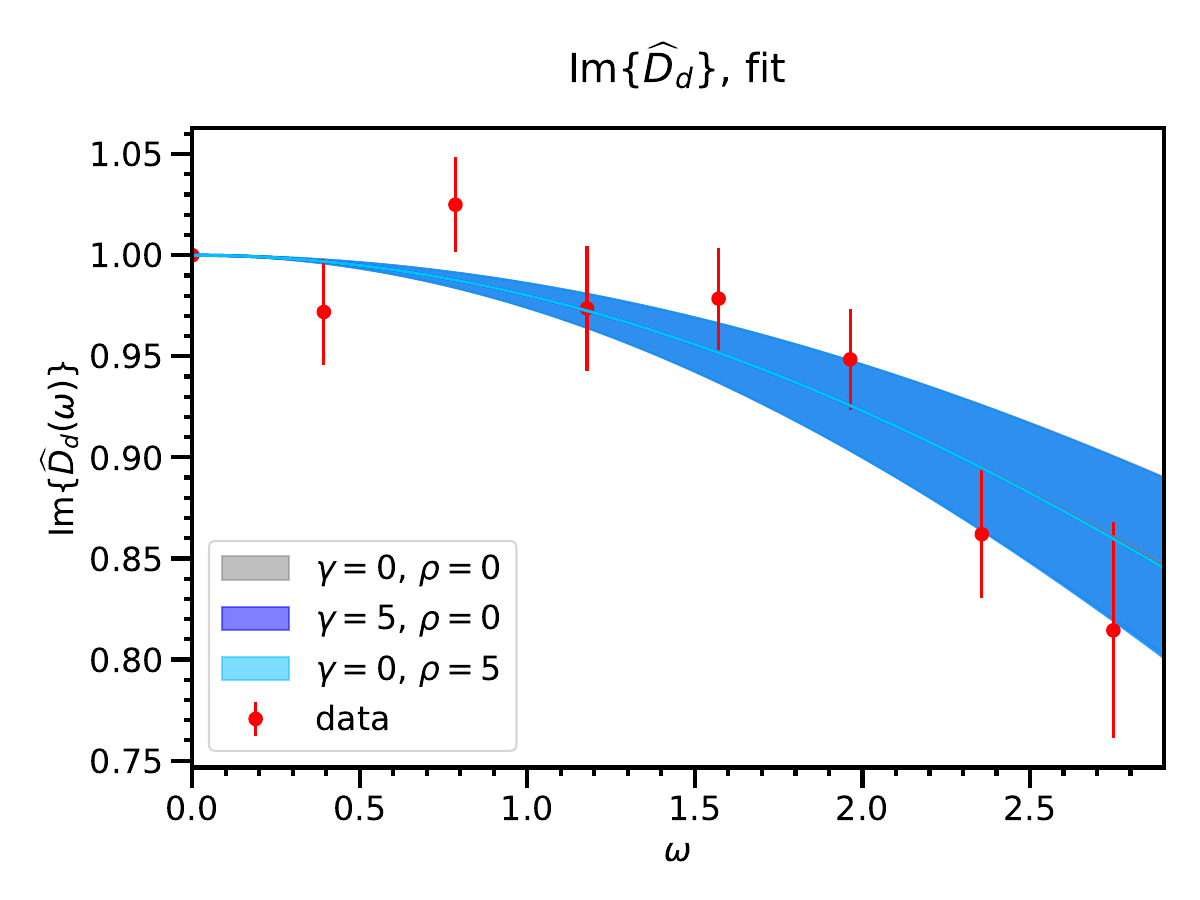}}
\caption{Fit results for $\mathrm{Im}\{\widehat{D}_{q}(\omega)\}$ for quark flavors $u$ (a) and $d$ (b) using the fit ansatz derived from \eqref{eq:im-D-integ} and \eqref{eq:pdf-ansatz}. \label{fig:imDhatomega-fit-grcomp}}
\end{figure}
In \fig\ref{fig:imDhatomega-fit-grcomp} we show the fit curves compared to the data points for flavor $u$ (a) and flavor $d$ (b). It turns out that there are almost no visible differences between the different combinations of $\gamma$ and $\rho$. This observation is in agreement with the fact that the value of $\chi^2/\mathrm{d.o.f.}$ varies only marginally between the considered fits for a given channel. Although there is a slight preference for larger values of $\gamma$ or $\rho$, the results obtained for different combinations of $\gamma$ and $\rho$ serve as input to estimate the systematic uncertainties introduced by the limitation in Ioffe time rather than using them to fine-tune the fit. 

\begin{figure}
\vspace*{0.5cm}
\subfigure[result for $f^p_{u,v}$\label{fig:pdf-u-fit-ab-var-comp}]
{\includegraphics[scale=0.43, clip, trim=0 0 0 1.6cm]{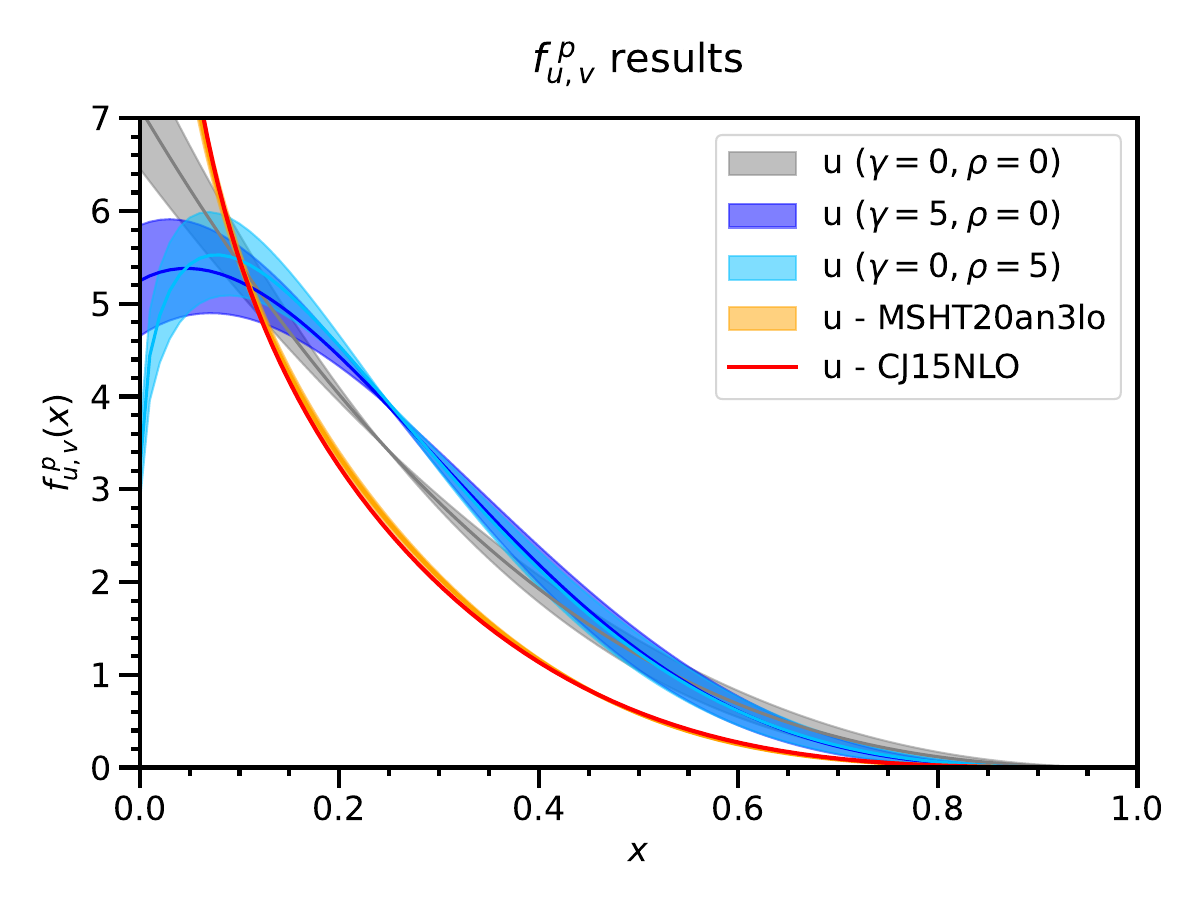}}
\vspace*{0.5cm}
\subfigure[result for $f^p_{d,v}$\label{fig:pdf-d-fit-ab-var-comp}]
{\includegraphics[scale=0.43, clip, trim=0 0 0 1.6cm]{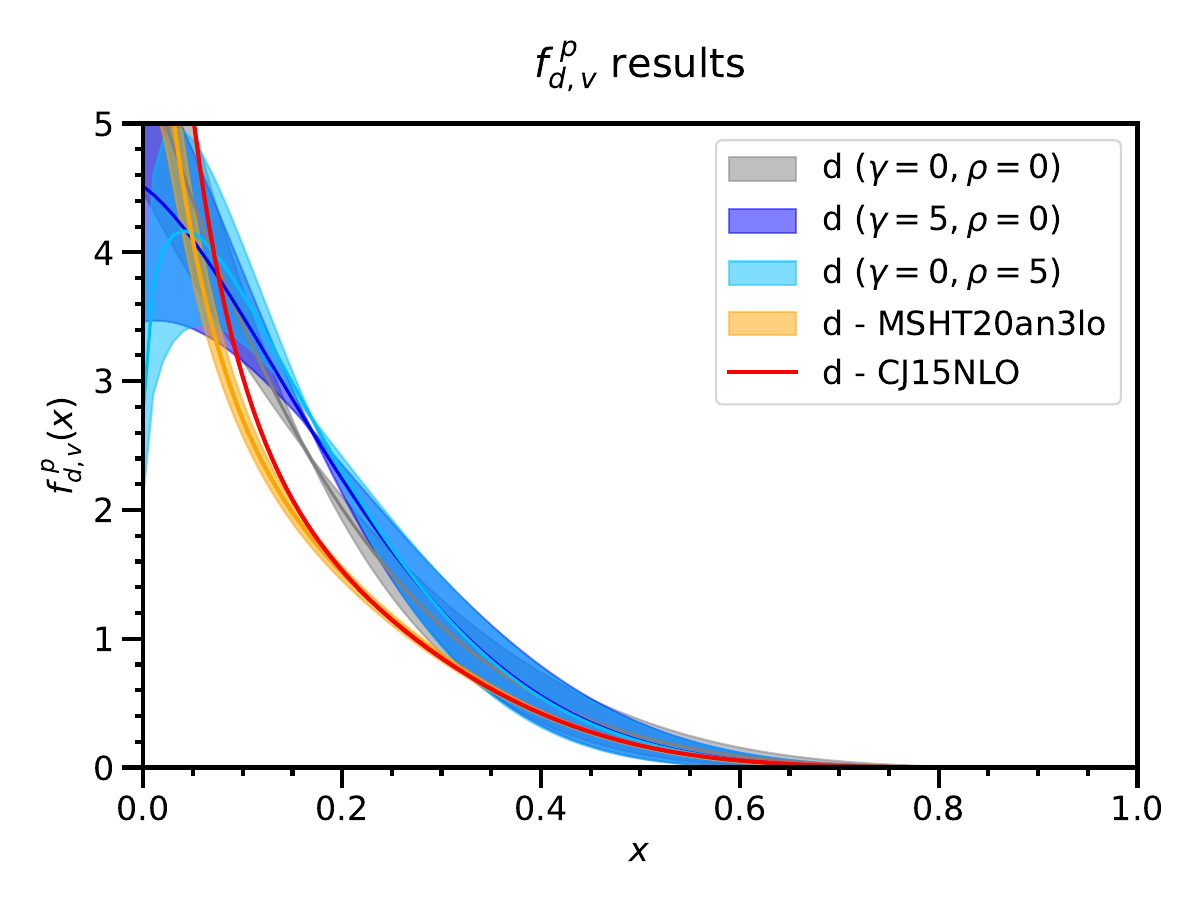}}
\caption{Results for the proton PDFs $f^p_{u,v}$ (a) and $f^p_{d,v}$ (b) plotted as a function of the quark momentum fraction $x$ for two different fits. The results are compared to experimental data (red and orange) \cite{Accardi:2016qay,McGowan:2022nag} for an evolution scale of $\mu = 2~\mathrm{GeV}$.\label{fig:pdf-fit-ab-var-comp}}
\end{figure}

Let us finally have a look at the result for the $x$ dependence of the PDFs, which we obtain by inserting the results for the fit parameters in the ansatz \eqref{eq:pdf-ansatz}. This is plotted in \fig\ref{fig:pdf-fit-ab-var-comp} for $f^p_{u,v}$ (a) and $f^p_{d,v}$ (b). It is observed that the results for different choices of $\gamma$ or $\rho$ are consistent within the error for large values of $x$. For smaller $x$, differences become larger, especially in the region $x<0.1$, which is a consequence of the limitation of the Ioffe time region. We observe that the curves for the $d$ quark go faster to zero than those for the $u$ quark, which is a consequence arising from a consistently larger value of $\beta$ for the $d$ quark; see \tab\ref{tab:omega-fit-results}. A faster decrease in the case of the $d$ quark is also a well-known result in PDF phenomenology.

In \fig\ref{fig:pdf-fit-ab-var-comp} we also compare with experimental data for the valence quark distributions \cite{Accardi:2016qay,McGowan:2022nag} (red and orange), which are evolved to an evolution scale of $\mu = 2~\mathrm{GeV}$ Discrepancies between the lattice results and the experimental data are present for both flavors. Whereas they are moderate for the $d$ quark (agreement within the statistical error in a wide range of $x$), the differences are more pronounced in the case of the $u$ quark. Notice that larger differences to the experimental results for the $u$ quark have been already observed on the level of the function $\widehat{D}$ in Ioffe time space; see \fig\ref{fig:imDhat-omega}.

The discrepancies between lattice results and experimental data can have several sources. First of all, our calculation of the matching coefficients was carried out at leading-order perturbation theory. The slight $y$ dependence observed in the data for $\mvec{p} = \mvec{0}$ (see \fig\ref{fig:imD-ydep}) may be a hint to higher-order contributions. However, for $\mvec{p} \neq \mvec{0}$, where the statistical errors are larger, these effects are likely to play only a minor role. Another source is give by higher-twist contributions, which are not contained in our ansatz. In analog calculations for the pion by \cite{Sufian:2019bol,Sufian:2020vzb}, a heavy intermediate quark was used in order to mitigate their effects. In our present simulation, all quarks have the same (light) mass. 

The systematic uncertainty of our analysis is to a large extent governed by the restriction of the accessible range in Ioffe time $\omega$. In particular, this affects our results obtained for the small-$x$ region. The limitation of the accessible $\omega$ region results from the restriction of the operator distance in order to keep higher-twist contributions small. Again, understanding the effect of higher-twist contributions might help to improve the situation. On the other hand, the accessible Ioffe time range can be increased by larger nucleon momenta. Moreover, we want to emphasize that our analysis is carried out only for one lattice spacing and unphysical quark masses. The impact of discretization errors and a potential dependence on the quark masses still has to be investigated.

\section{\label{sec:concl}Conclusions}

We adapted the LCS framework developed in \cite{Ma:2017pxb} to the case of valence quark PDFs in the nucleon. The corresponding two-current matrix elements can be evaluated on the lattice for purely spatial current separations. In the present study we reuse the four-point functions produced in the context of another project \cite{Bali:2021gel}. For valence quark PDFs, only the data for the $C_2$ contraction are needed. Lattice artifacts have been reduced by considering a suitable combination of currents as well as a simple tree-level improvement of the Wilson propagator. 

From the two-current matrix element, we extracted the Lorentz invariant function $D_q$ for the quark flavors $u$ and $d$, which, at leading order, is the Fourier transform of the PDF. The data for $\omega = 0$ were observed to be consistent with the quark number sum rule. The results we obtained for the normalized invariant function $\widehat{D}_q$ agrees very well with the experimental data up to $\omega = 2$. Beyond that point, deviations start to be visible, which are stronger in the $u$-quark case. In order to extract the $x$ dependence of PDFs, we use a conventional ansatz for the functional form of PDFs; see \eqref{eq:pdf-ansatz}. Its Fourier transform serves as ansatz to perform a fit to our lattice data. We performed several fits varying the fixed values of the parameters $\rho$ and $\gamma$. The corresponding values for $\chi^2/\mathrm{d.o.f.}$ are comparable. From these fits we obtain results for the $x$ dependence of the valence quark PDFs with reasonable statistical error. We observe that for increasing $x$, the corresponding curve for the down quark approaches zero faster than in the case of the up quark, which is in agreement with PDF phenomenology. Direct comparisons with experimental PDF data reveal differences. These are moderate in the case of the $d$ quark, whereas they are more pronounced for the $u$-quark PDF.

There is room for several improvements. For instance, the perturbative determination of the matching coefficient can be extended by including next-to-leading-order contributions. Furthermore, it is advisable to understand the role of higher-twist contributions, which have been neglected so far. Moreover, considering simulations with higher nucleon momenta would increase the accessible Ioffe time range. The limitation there is the main source for systematic uncertainties of our PDF results in the small-$x$ region. Higher momenta would also be of interest in the context of the DPD project, within which our four-point functions have been generated. Another source of uncertainties is the treatment of the inverse problem, where we currently use a fit inspired by common PDF models. In order to deal more appropriately with the inverse problem, considering more sophisticated methods like a Bayesian analysis might be advisable. Moreover, we have to investigate potential discretization errors and effects caused by the unphysical quark masses employed in the current analysis. Simulations for further ensembles closer to the physical point are currently in progress.


\begin{acknowledgments}
We gratefully thank Markus Diehl, Gunnar Bali, and Raza Sabbir Sufian for fruitful discussions. Moreover, we thankfully acknowledge the effort by the CLS Collaboration of generating the $n_f = 2+1$ gauge ensembles. The lattice simulations have been performed on the SFB TRR55 QPACE3. The project leading to this publication received funding from the Excellence Initiative of Aix-Marseille University - A*MIDEX, a French “Investissements d’Avenir” programme, AMX-18-ACE-005. Our DPD effort is supported by DFG grant SCHA 458/23.
\end{acknowledgments}



\bibliography{nucl-xpdf-paper}

\end{document}